\documentclass[11pt]{article}
\usepackage{amsmath}

\title{
\hfill
\parbox{3cm}{\normalsize DPNU-02-23\\
{\tt hep-lat/0207013}}\\
\vspace{0.5cm}
Nicolai mapping vs. exact chiral symmetry\\
on the lattice
} 
\author{
Yoshio Kikukawa\thanks{e-mail address:
kikukawa@eken.phys.nagoya-u.ac.jp} 
and 
Yoichi Nakayama\thanks{e-mail address:
yoichi@eken.phys.nagoya-u.ac.jp} 
\\
\\
{\normalsize\em Department of Physics, Nagoya University 
}\\
{\normalsize\em Nagoya 464-8602, Japan}
\\
\\
\date{\normalsize July, 2002}
}
\begin{document}

\maketitle

\begin{abstract}
Two-dimensional N=2 Wess-Zumino model is constructed on the lattice
through Nicolai mapping with Ginsparg-Wilson fermion. 
The Nicolai mapping requires a certain would-be surface term in the
bosonic action which ensures the vacuum energy cancellation even on
the lattice, but inevitably breaks chiral symmetry. With the
Ginsparg-Wilson fermion, the holomorphic structure of the 
would-be surface term is maintained, leaving a discrete subgroup 
of the exact chiral symmetry intact for a monomial scalar potential. 
By this feature both boson and fermion can be kept massless 
on the lattice without any fine-tuning.
\end{abstract}

\newpage
\section{Introduction}
The recent re-discovery of the Ginsparg-Wilson 
relation\cite{Ginsparg-Wilson,PH-fp-action,HN-chiralsym} and the 
realization of exact chiral symmetry on the lattice \cite{ML-chiralsym}
are interesting developments from the point of view of the constructive
approach to quantum field theory.\footnote{
Fujikawa has proposed a new class of Dirac operators
by the algebraic extension of the Ginsparg-Wilson
relation\cite{Fujikawa-algebraic-extension}.}
It is  a challenge to extend this 
idea to other aspects of quantum field theory. The construction of 
supersymmetric theories is one possibility in this direction,
although it has been known to be difficult because of 
the lack of infinitesimal translation invariance on the lattice and
the breakdown of the Leibniz 
rule\cite{staggered-susy,slac-susy,nojiri,curci-veneziano,wilson-susy,taniguchi-WT-Wilson}. 
Based on domain wall fermion\cite{dwf,dwf-vector}, 
overlap formalism\cite{overlap} and the Ginsparg-Wilson 
relation, there are several 
attempts\cite{nishimura-SYM,neuberger-SYM,aoyama-kikukawa,kaplan-shmaltz,
bietenholz-susy,so-ukita-susy,KF-MI-wz,Fujikawa-SUSY-Leibniz} so far.

Despite the difficulties, 
two-dimensional N=2 Wess-Zumino model has been constructed 
successfully 
based on 
the Nicolai mapping\cite{Nicolai-mapping}
in the Hamiltonian 
formalism by Cecotti and Girardello\cite{spacial-lattice-nicolai} 
and on a Euclidean lattice by Sakai and Sakamoto\cite{NS-MS-nicolai},
respectively.
The Nicolai mapping is the transformation of the bosonic field variables
to the gaussian stochastic variables whose Jacobian just reproduces
the functional determinant of the fermions in the model.
The Euclidean lattice version of the Nicolai 
mapping\footnote{
The Nicolai mapping on the spacial lattice 
in the Hamiltonian formalism was first constructed by 
Cecotti and Girardello in \cite{spacial-lattice-nicolai}. 
The Nicolai mapping on the two-dimensional Euclidean lattice was
obtained 
by Sakai and Sakamoto in \cite{NS-MS-nicolai}.}
produces a certain would-be surface term in the bosonic 
action and ensures the vacuum energy
cancellation even on the lattice!
Moreover, one special combination out of four supersymmetries of 
the N=2 model is manifest in the lattice action.\footnote{
The lattice model with certain fermionic symmetry
has recently been proposed by Itoh, Kato, Sawanaka, So and
Ukita\cite{so-kato-susy}.}
\footnote{
In the same spirit, but in a quite new approach, 
the construction of super Yang-Mills theory on the 
spacial lattice has recently been proposed by Kaplan, Katz and
Unsal\cite{kaplan-susy-spatial-lattice}.}

In this construction, however, 
the remaining three supersymmetries cannot be maintained.
As clarified by Catterall and Karamov\cite{SC-SK-susy},
the four different supersymmetries in the
original model can be associated with the four different methods to 
construct the Nicolai mapping. The resulted four different would-be 
surface terms reduce to surface terms
in the continuum limit through the Leibniz rule, 
and then the four supersymmetries are realized at the same time.
But at finite lattice spacing they define 
four different lattice models and in each model only one supersymmetry 
is realized.

Another unsatisfactory feature of the above construction is that chiral
symmetry of the original model is not maintained and a fine-tuning is
required to keep the degenerate boson and fermion light or massless.
This is partly because the fermion theory obtained through the lattice
Nicolai mapping turns out to be the Wilson-Dirac fermion. More
seriously, the would-be surface term required in the bosonic action
breaks chiral symmetry explicitly.

The purpose of this letter is to construct
two-dimensional N=2 Wess-Zumino model with the Ginsparg-Wilson fermion
and examine the above problems.
We construct the lattice Nicolai mapping so that its Jacobian
reproduces the functional determinant of the Ginsparg-Wilson
fermion possessing Yukawa coupling with the exact chiral symmetry.
We will see that the use of the Ginsparg-Wilson fermion
improves the holomorphic structure of the would-be surface term.
Although it still breaks chiral symmetry explicitly in general,
but for monomial scalar potentials, 
\begin{equation}
  W[\phi]=\lambda \phi^n, \quad n=3,4,5,\cdots
\end{equation}
it leaves a discrete subgroup of exact chiral symmetry intact 
and both boson and fermion can be kept massless on the lattice 
without any fine-tuning.

We will also discuss how the asymmetric treatment between 
the field and antifield of the Ginsparg-Wilson fermion
affects the structure of the Nicolai mapping.
Actually, because of the asymmetric treatment,
the Cauchy-Riemann condition can be satisfied for only two cases
out of four possible Nicolai mappings
discussed by Catterall and Karamov\cite{SC-SK-susy}.

\section{Two-dimensional N=2 Wess-Zumino model \\
-- Nicolai mapping and supersymmetry}

The action of the two-dimensional N=2 Wess-Zumino model 
in the continuum limit is give by
\begin{align}
 & S = S_B + S_F , \\
 & S_B  = \int {\rm d}^2x \, {\cal L}_B(x) = 
\int  {\rm d}^2x \left\{ 
  \partial_\mu \phi^* \partial_\mu \phi +
  W^*{}' W'
 \right\} , \\
 & S_F = \int {\rm d}^2x \, {\cal L}_F(x) = \int  {\rm d}^2x \left\{ 
  \bar\psi \gamma_\mu \partial_\mu \psi +
  \bar \psi W'' \frac{1+\gamma_3}{2} \psi +
  \bar \psi W^*{}'' \frac{1-\gamma_3}{2} \psi 
 \right\} .
\end{align}
This action is invariant under four independent supersymmetry 
transformations associated with four independent real grassmann 
parameters. The Lagrangian is invariant up to terms
which can be rewritten into a total divergence through the Leibniz rule. 
This property of the supersymmetry transformations immediately
causes a trouble on the lattice, because the Leibniz rule
does not hold for the field products of more than quadratic orders. 

This model, however, possesses the so-called Nicolai mapping:
\begin{align}
&  M(x) =  -\partial_1 A(x) - \partial_2 B(x) + U(x), \nonumber\\
&  N(x) =  -\partial_2 A(x) + \partial_1 B(x) + V(x),
\end{align}
where $A,B$ and $U,V$ are real and imaginary parts of $\phi$ and $W'$,
respectively, 
\begin{equation}
 \phi = \sqrt{\frac{1}{2}}(A+iB)
 ,\quad
 W'=\sqrt{\frac{1}{2}}(U+iV) .
\end{equation}
The Jacobian of this transformation of the bosonic field variables 
just coincides with the functional determinant of the fermion,
\begin{equation}
\det
\left(\begin{matrix}
  \frac{\partial M}{\partial A} &
  \frac{\partial N}{\partial A}\\
  \frac{\partial M}{\partial B} &
  \frac{\partial N}{\partial B}
\end{matrix}\right)
  = 
\det \left\{
\gamma_\mu \partial_\mu +
  W'' \frac{1+\gamma_3}{2} + W^*{}'' \frac{1-\gamma_3}{2} 
 \right\}, 
\end{equation}
while the gaussian weight for $M(x)$ and $N(x)$ reproduces
the bosonic part of the Lagrangian, ${\cal L}_B(x)$, 
\begin{equation}
\frac{1}{2}\{M(x)^2+ N(x)^2\}
= 
\partial_\mu \phi^* \partial_\mu \phi +  W^*{}' W' 
+ 
W' \partial_{\bar z} \phi +  W^*{}'\partial_z
\phi^* \equiv {\cal L}^\prime_B(x)
\end{equation}
up to the surface terms, 
$W' \partial_{\bar z} \phi +  W^*{}'\partial_z\phi^*
= \partial_{\bar z} W + \partial_z W^* $.
The gaussian path-integral of
$M(x)$ and $N(x)$ can reproduce the partition function of 
the original model.

From the structure of the above Nicolai mapping, it follows that 
the action is invariant under the following fermionic 
transformation\cite{Parisi-Sourlas}, 
\begin{equation}
 \delta A = \bar \psi_1 \xi ,\quad
 \delta B = -i \bar \psi_2 \xi
 \label{eq:susy-transf-scalar}
\end{equation}
\begin{equation}
 \delta \psi_1 = -\xi M , \quad
 \delta \psi_2 = i \xi N 
 \label{eq:susy-transf-fermion}
\end{equation}
\begin{equation}
 \delta \bar\psi_1 = 0 ,\quad
 \delta \bar\psi_2 = 0
 \label{eq:susy-transf-anti-fermion}
\end{equation}
where $\xi$ is a one-component grassmann parameter and
\begin{equation}
\psi = 
 \left(\begin{matrix}
\psi_1 \\
\psi_2
\end{matrix}
\right) \quad ,\quad
\bar\psi = \left(
\bar\psi_1 , \bar\psi_2
\right) .
\end{equation}
This transformation is a certain combination of the supersymmetry
transformation of the N=2 model, which has a special feature:
the total Lagrangian which includes the extra surface terms
required by the Nicolai mapping, 
${\cal L}^\prime(x) = {\cal L}_B^\prime(x)+{\cal L}_F(x)$, 
is exactly invariant without leaving any surface term.
See appendix for detail. 
Therefore, 
this special supersymmetry has a fair chance to be realized on the lattice.

In fact, as shown by Sakai and Sakamoto\cite{NS-MS-nicolai}, 
the Nicolai mapping can be constructed successfully on the 
two-dimensional Euclidean lattice. Their result reads
\begin{align}
&  M(x) = (- \nabla_1^S - \nabla_1^A -\nabla_2^A )A (x)
          - \nabla_2^S B (x) + U(x),
\nonumber\\
&  N(x) = - \nabla_2^S (x)
          +(\nabla_1^S - \nabla_1^A -\nabla_2^A)B(x) + V(x),
\end{align}
where $\nabla^{A,S}$ are defined by forward and backward
differentials as
\begin{equation}
 \nabla_j^S = \frac12 \left(\nabla_j^+ + \nabla_j^- \right) \ , \
\nabla_j^A = \frac12 \left( \nabla_j^+ - \nabla_j^- \right).
\end{equation}
The Jacobian of this lattice Nicolai mapping reproduces
the functional determinant of the Wilson-Dirac fermion
with the Yukawa coupling 
\begin{equation}
\det
\left(\begin{matrix}
  \frac{\partial M}{\partial A} &
  \frac{\partial N}{\partial A}\\
  \frac{\partial M}{\partial B} &
  \frac{\partial N}{\partial B}
\end{matrix}\right)
  = 
\det \left\{
\sum_\mu \left( \gamma_\mu \nabla^S_\mu - \nabla^A_\mu \right)
 +
  W'' \frac{1+\gamma_3}{2} + W^*{}'' \frac{1-\gamma_3}{2} 
 \right\}, 
\end{equation}
while the bosonic action determined by the lattice Nicolai 
mapping contains the following ``would-be surface terms'',
\begin{align}
 &\phi (\nabla_1^S - i\nabla_2^S)W'
 + \phi^* (\nabla_1^S + i \nabla_2^S) W^*{}'
 \nonumber\\
 &- \phi (\nabla_1^A + \nabla_2^A) W^*{}'
 - \phi^* (\nabla_1^A + \nabla_2^A) W'
 \label{eq:wilson-would-be-surface-term}
\end{align}
By virtue of these terms, the vacuum energy cancellation holds on 
the lattice. Moreover, the total action 
possesses a supersymmetry under the same transformation as 
Eqs.~(\ref{eq:susy-transf-scalar}), (\ref{eq:susy-transf-fermion})
and (\ref{eq:susy-transf-anti-fermion}).

\section{Nicolai mapping with Ginsparg-Wilson fermion}
\label{sec:nicolai-for-GW-fermion}

Now we construct the two-dimensional N=2 Wess-Zumino model 
with the Ginsparg-Wilson fermion,
relying on the existence of the Nicolai
mapping as the guiding principle to maintain supersymmetry as in
\cite{NS-MS-nicolai}. 
Our strategy is as follows. First we fix the fermionic part 
of the action so that the Yukawa coupling possesses the exact chiral 
symmetry based on the Ginsparg-Wilson relation. Then we construct the Nicolai 
mapping so that its Jacobian reproduces the functional determinant
of the Ginsparg-Wilson fermion with the Yukawa coupling.
Finally, the bosonic part of the action is determined so that 
it coincides with the gaussian weight for the Nicolai-mapped 
bosonic variables.

We take the following fermionic action: 
\begin{align}
 S_F &= \sum_x \bar \psi (D+F)\psi \nonumber\\
  &=
  \sum_{x,y}
  \bar\psi (x) \left( D +
  \frac{1+\gamma_3}{2} W'' \frac{1+\hat\gamma_3}{2} +
  \frac{1-\gamma_3}{2} W^{*}{}'' \frac{1-\hat\gamma_3}{2}
  \right)_{x,y} \psi (y).
 \label{eq:fermionic-action}
\end{align}
where $D$ is a lattice Dirac operator which satisfies the Ginsparg-Wilson
relation,
\begin{equation}
 D \hat\gamma_3 + \gamma_3 D = 0
  ,\quad
 \hat\gamma_3 = \gamma_3 (1-aD)
 .
\end{equation}
As an explicit example, we adopt the overlap Dirac
operator given by Neuberger \cite{HN-chiralsym}.
\begin{equation}
 D = \left(
      \begin{matrix} 
       T+S_1 & iS_2 \\
       -iS_2 & T - S_1
      \end{matrix}
     \right)
 , \label{eq:overlap-D}
\end{equation}
where $T$, $S_1$, $S_2$ are defined as
\begin{equation}
 T=\frac{1}{a}\left(
 1- \frac{1}{\sqrt{X^\dagger X}}
\right)
-
\frac{\nabla_1^A+\nabla_2^A}{\sqrt{X^\dagger X}} 
={}^tT
,
\end{equation}
\begin{equation}
 S_j= \frac{\nabla_j^S}{\sqrt{X^\dagger X}} = - {}^tS_j
 ,\quad j=1,2
\end{equation}
\begin{equation}
 X = 1-aD_W
 .
\end{equation}
In this notation, the Ginsparg-Wilson relation can be written as
\begin{equation}
 a (T^2 - S_1^2 -S_2^2) = 2 T
 .
 \label{eq:ginsparg-wilson-relation}
\end{equation}
By construction, 
the fermionic part of the action (\ref{eq:fermionic-action}) is
invariant under lattice chiral rotation \cite{ML-chiralsym}
\begin{gather}
 \psi \rightarrow \exp \left(i\theta \hat\gamma_3\right)\psi, \quad
 \bar\psi \rightarrow \bar\psi \exp \left(i\theta \gamma_3\right)
 , \nonumber\\
 W'' \rightarrow W'' \exp \left(2i\theta \right), \quad
 W^{*}{}'' \rightarrow W^{*}{}'' \exp \left(-2i\theta \right)
 . \label{eq:chiral-rotation}
\end{gather}

By inserting the Dirac operator (\ref{eq:overlap-D})
into (\ref{eq:fermionic-action}), we obtain
\begin{align}
 & D + F = 
 \left(
\begin{matrix} 
T+S_1& iS_2\\
-iS_2& T - S_1
\end{matrix}
\right) + \nonumber\\
 & \left(
\begin{matrix} 
\frac{\partial U}{\partial A}\left(1-\frac{a}{2}(T+S_1)\right)
 -\frac{\partial V}{\partial A}\frac{a}{2} S_2&
i\left\{ \frac{\partial V}{\partial A}\left(1-\frac{a}{2}(T-S_1)\right)
-\frac{\partial U}{\partial A}\frac{a}{2}S_2\right\}\\
-i\left\{\frac{\partial U}{\partial
 B}\left(1-\frac{a}{2}(T+S_1)\right)
-\frac{\partial V}{\partial B}\frac{a}{2}S_2\right\}
&
\frac{\partial V}{\partial B}\left(1-\frac{a}{2}(T-
 S_1)\right)
-\frac{\partial U}{\partial B}\frac{a}{2}S_2
\end{matrix}\right)
 \label{eq:fermionic-matrix}
\end{align}
where $A,B,U,V$ are real and imaginary parts of $\phi,W'$
\begin{equation}
 \phi = \sqrt{\frac{1}{2}}(A+iB)
 ,\quad
 W'=\sqrt{\frac{1}{2}}(U+iV)
 .
\end{equation}
Then the Nicolai mapping should solve the differential equation
\begin{align}
D + F =\left(\begin{matrix}
  \frac{\partial M}{\partial A} &
  i\frac{\partial N}{\partial A}\\
  -i\frac{\partial M}{\partial B} &
  \frac{\partial N}{\partial B}
\end{matrix}\right)
 .
 \label{eq:differential-equation-nicolai}
\end{align}
We can find a solution to this equation as follows:
\begin{equation}
 M = A(T+S_1) + BS_2 + U\left(1-\frac{a}{2}(T+S_1)\right)-V\frac{a}{2}S_2 ,
\label{eq:GW-fermion-nicolai-mapping-M}
\end{equation}
\begin{equation}
 N = AS_2 +B(T-S_1) + V\left(1-\frac{a}{2}(T-S_1)\right)-U\frac{a}{2}S_2 ,
\label{eq:GW-fermion-nicolai-mapping-N}
\end{equation}
where $M,N,A,B,U,V$ are functions of $x$ and difference operators
$T,S_1,S_2$ are multiplied from the right.
As to other possible solutions, we will discuss later.

We now evaluate the bosonic part of the action implied by the above 
Nicolai mapping,
\begin{equation}
 S_B = \frac{1}{2}\sum_x \left\{ M^2 + N^2\right\} .
\end{equation}
The Ginsparg-Wilson relation plays an important role through the
calculation: as an illustrative example, we show $A\times U$ term and
$B\times V$ term,
\begin{align}
 & A \left( S_1 + T -\frac{a}{2}(T^2-S_1^2-S_2^2)\right) U
 + B \left( -S_1 + T -\frac{a}{2}(T^2-S_1^2-S_2^2)\right) V
\nonumber\\
&= \phi^* \left( T-\frac{a}{2}(T^2-S_1^2-S_2^2) \right) W'
  + \phi \left( T-\frac{a}{2}(T^2-S_1^2-S_2^2) \right) W^*{}'
\nonumber\\
 & + \phi S_1 W' + \phi^* S_1 W^*{}'
.
\label{eq:illustrative-example}
\end{align}
Here we note that the combination $T-\frac{a}{2}(T^2-S_1^2-S_2^2)$ 
is equal to zero
by (\ref{eq:ginsparg-wilson-relation}). 
We finally obtain the bosonic part of the action as
\begin{align}
 S_B 
 = \ &
 \sum_x \Bigl\{
 \phi^* \Delta \phi +
 W^*{}'(1-\frac{a^2}{4}\Delta ) W' \nonumber\\
 & +  W' (-S_1 + iS_2)\phi 
 + W^*{}' (-S_1 - iS_2)\phi^* \Bigr\}
 \label{eq:bosonic-action}
\end{align}
where $\Delta$ is defined by $D^\dagger D = \Delta \cdot {\mathbf 1}$ and
$\Delta = (T^2 - S_1^2 -S_2^2) = 2 T/a$.

Thanks to the existence of the Nicolai mapping,  
(\ref{eq:GW-fermion-nicolai-mapping-M}) and 
(\ref{eq:GW-fermion-nicolai-mapping-N}), 
it is ensured that 
all the nice features of the construction by Sakai and
Sakamoto \cite{NS-MS-nicolai} are maintained in our construction. 
The total action $S=S_B + S_F$ given by
(\ref{eq:bosonic-action}) and (\ref{eq:fermionic-action}), 
possesses a supersymmetry under the transformation
\begin{equation}
 \delta A = \bar \psi_1 \xi ,\quad
 \delta B = -i \bar \psi_2 \xi
\end{equation}
\begin{equation}
 \delta \psi_1 = -\xi M , \quad
 \delta \psi_2 = i \xi N 
\end{equation}
\begin{equation}
 \delta \bar\psi_1 = 0 ,\quad
 \delta \bar\psi_2 = 0
\end{equation}
where $\xi$ is a one-component grassmann parameter and
\begin{equation}
\psi = 
 \left(\begin{matrix}
\psi_1 \\
\psi_2
\end{matrix}
\right) \quad ,\quad
\bar\psi = \left(
\bar\psi_1 , \bar\psi_2
\right).
\end{equation}
The vacuum energy cancellation also holds 
even at the finite lattice spacing.
(One may verify through
explicit calculations that 
the vacuum energy is canceled exactly in any orders of 
the lattice perturbation theory.)  

\section{Chiral symmetry in the supersymmetric action}

Now let us examine the chiral properties of the lattice action 
of the two-dimensional N=2 Wess-Zumino model 
obtained in the previous section. The fermionic part of 
the action, (\ref{eq:fermionic-action}), respects the exact 
chiral symmetry on the lattice by our construction.
Then the question is the chiral properties of the bosonic
part of the action, (\ref{eq:bosonic-action}).

First of all, the bosonic part of the action, (\ref{eq:bosonic-action}), 
should be compared with the counterpart in the construction by Sakai and
Sakamoto, (\ref{eq:wilson-would-be-surface-term}), or
the equation (3.6) in \cite{NS-MS-nicolai}.
An important difference is in that 
the terms with the structures, $W'\times \phi^*$ and $W^*{}'\times \phi$, 
do not appear in (\ref{eq:bosonic-action}), 
and this implies that the holomorphic structure of the
would-be surface terms is maintained just as in the continuum theory. 
As we have seen explicitly in (\ref{eq:illustrative-example}),
these terms vanish identically by virtue of the Ginsparg-Wilson relation. 
Thus the use of the Ginsparg-Wilson fermion can improve the holomorphic 
structure of the would-be surface term.

The would-be surface terms in (\ref{eq:bosonic-action}) 
still break the exact chiral symmetry on the lattice explicitly.
They cannot be eliminated, because these terms are playing a crucial role 
in order to maintain the supersymmetry of the action.
Therefore the breakdown of the 
exact chiral symmetry on the lattice seems inevitable.

Thanks to the improved holomorphic structure, however,
if one assumes that the superpotential is a monomial
\begin{equation}
 W(\phi) = \lambda \phi ^ n ,\quad n =3,4,5,\cdots
  ,
\end{equation}
then the total action is invariant under
the discrete chiral rotation 
with the angle $\theta = \pi k /n$ for arbitrary integer $k$.
By this remaining discrete exact chiral symmetry, 
both boson and fermion can be kept massless 
on the lattice without any fine-tuning.
We would have the same situation in the continuum theory 
if we keep the total divergence term implied by the Nicolai mapping
in the action so that an exact supersymmetry is maintained 
at the Lagrangian level. 
So, we think, it is not quite a lattice artifact.

It is not difficult to prove 
in any order of the lattice perturbation expansion 
that 
the fermion mass term would not be produced
in this lattice model with a monomial potential. 
The possible coupling terms appear in the
following combinations
\begin{equation}
 \phi^{n-1}\phi^{*n-1}, \  \phi^{n} ,  \ \phi^{*n} ,
 \bar\psi_L \phi^{n-2} \psi_R , \ \bar\psi_R \phi^{*n-2} \psi_L
\end{equation}
where we omit derivatives and proportional factors. In
perturbation expansion, we should consider all possible diagrams 
produced by the product of those couplings. 
The mass term must have the external legs $\bar\psi_L$-$\psi_R$ (or
$\bar\psi_R$-$\psi_L$), while the $n-2$ legs of scalar field coming from
the combination
$(\bar\psi_L \phi^{n-2} \psi_R)^{(l+1)} (\bar\psi_R \phi^{*n-2} \psi_L)^l$
($l=0,1,2,\cdots$) cannot be closed by $-n$ legs coming from $\phi^{*n}$
or by any other product of the
interaction terms~\footnote{$-j n = n -2$ cannot be satisfied by any
integer $j$ for $n=3,4,5,\cdots$.}. Therefore we can conclude that the 
fermion mass term would not be generated in our model.
Then the supersymmetry implies that the boson would not acquire mass, neither.

Here we should emphasize that the same 
result cannot be obtained in the case of the Wilson fermion,
because there are no mechanism to suppress non-holomorphic scalar
self-interaction.

\section{Solubility of Nicolai mappings}

Two-dimensional N=2 Wess-Zumino model is invariant under
four supersymmetry transformations which can be related to four types
of the Nicolai mappings as clarified in \cite{SC-SK-susy}. In the case
with Wilson-Dirac fermions, we can actually obtain all the four mappings.

In the case with Ginsparg-Wilson fermions, however,
the situation differs due to the asymmetric choice
of chiral projectors (\ref{eq:fermionic-action}).
The four differential equations corresponding to
the four Nicolai mappings are given by
\begin{align}
 D+F &
 =\left(\begin{matrix}
  \frac{\partial M}{\partial A} &
  i\frac{\partial N}{\partial A} \\
  -i\frac{\partial M}{\partial B} &
  \frac{\partial N}{\partial B}
\end{matrix}\right) ,
\label{eq:first-Nicolai}\\
 D+F &
 =\left(\begin{matrix}
  \frac{\partial M}{\partial B} &
  -i\frac{\partial N}{\partial B} \\
  i\frac{\partial M}{\partial A} &
  \frac{\partial N}{\partial A}
\end{matrix}\right) ,
\label{eq:second-Nicolai}\\
 D+F &
 =\left(\begin{matrix}
  \frac{\partial M}{\partial A} &
  -i\frac{\partial M}{\partial B} \\
  i\frac{\partial N}{\partial A} &
  \frac{\partial N}{\partial B}
\end{matrix}\right) ,
\label{eq:third-Nicolai}\\
 D+F &
 =\left(\begin{matrix}
  \frac{\partial M}{\partial B} &
  i\frac{\partial M}{\partial A} \\
  -i\frac{\partial N}{\partial B} &
  \frac{\partial N}{\partial A}
\end{matrix}\right).
\label{eq:fourth-Nicolai}
\end{align}
The solution of the first one (\ref{eq:first-Nicolai}) is
the solution given in section \ref{sec:nicolai-for-GW-fermion}.
The solution of the second one (\ref{eq:second-Nicolai})
is obtained in the similar manner using
$\frac{\partial U}{\partial A}=\frac{\partial V}{\partial B}$
and
$\frac{\partial U}{\partial B}=
- \frac{\partial V}{\partial A}$.\footnote{
The bosonic action given by the solution of
(\ref{eq:second-Nicolai})
has the form (\ref{eq:bosonic-action}) with the sign of $S_j$
reversed.}
However the rest two cases cannot be solved.
The Cauchy-Riemann condition, which is the necessary condition
for the solubility, does not hold for the latter two cases.
For example, the Cauchy-Riemann condition for
the third one (\ref{eq:third-Nicolai}) is evaluated as
\begin{equation}
 \frac{\partial}{\partial B} (D+F)_{11} -i \frac{\partial}{\partial A}
 (D+F)_{12}
 = - a 
 \left\{
 \frac{\partial^2 U}{\partial A \partial B} S_1 + 
 \frac{\partial^2 V}{\partial A \partial B} S_2
 \right\}
 \neq 0
 .
 \label{eq:cauchy-riemann-opposite}
\end{equation}
This violation of the Cauchy-Riemann condition 
is the consequence of the asymmetric choice of the chiral projectors.
Therefore the Nicolai mappings related to the other
two supersymmetries have no solutions.

If we perform singular change of the field variables as
\begin{equation}
 \psi' = (1-\frac{a}{2}D)\psi
  ,\quad
 \bar\psi' = \bar\psi(1-\frac{a}{2}D)^{-1} ,
 \label{eq:singular-mapping}
\end{equation}
then we can solve the
differential equations which correspond to 
(\ref{eq:third-Nicolai}) and
(\ref{eq:fourth-Nicolai}),
while the Cauchy-Riemann conditions for the equations
which correspond to
(\ref{eq:first-Nicolai}) and
(\ref{eq:second-Nicolai})
break down.

\section{Summary}

We have constructed two-dimensional N=2 Wess-Zumino model on the lattice
which possesses both the supersymmetry based on the Nicolai mapping
and the exact chiral symmetry based on the Ginsparg-Wilson relation. 
The Nicolai mapping ensures that the vacuum energy cancellation holds and 
boson and fermion are degenerate.
The use of the Ginsparg-Wilson fermion 
maintains the holomorphic structure of the would-be surface term, 
leaving a discrete subgroup of the exact chiral symmetry intact 
for a monomial scalar potential. 
Thus both boson and fermion can be kept massless on the lattice 
without any fine-tunning.

\section*{Acknowledgments}
Y.N. is supported in part by the Japan
Society for Promotion of Science under the Predoctoral
Research Program No.~12-4298.
Y.K. would like to thank T.~Onogi for valuable discussions.
Y.K. is supported in part by Grant-in-Aid 
for Scientific Research (\#12640262, \#14046207).

\appendix
\section{Nicolai mapping and supersymmetry}

In this appendix we examine the properties of the supersymmetry which
follows from the Nicolai mapping in the continuum theory.
Supercharges in two-dimensional N=2 theory are written as
\begin{equation}
 Q_+ = \frac{1}{\sqrt 2}\left(
 \frac{\partial}{\partial \bar \theta_-} - \theta_- \frac{\partial}{\partial z}
\right)
, \quad
 \overline Q_+ = \frac{1}{\sqrt 2}\left(
 \frac{\partial}{\partial \theta_-} - \bar \theta_- \frac{\partial}{\partial z}
\right),
\end{equation}
\begin{equation}
 Q_- = \frac{1}{\sqrt 2}\left(
 \frac{\partial}{\partial \theta_+} - \bar \theta_+
 \frac{\partial}{\partial \bar z}
\right)
, \quad
 \overline Q_- = \frac{1}{\sqrt 2}\left(
 \frac{\partial}{\partial \bar \theta_+} - \theta_+
 \frac{\partial}{\partial \bar z}
\right).
\end{equation}
These Qs satisfy following SUSY algebra
\begin{equation}
 \left\{ Q_+ , \overline Q_+ \right\} = - \frac{\partial}{\partial z} ,\quad
 \left\{ Q_- , \overline Q_- \right\} = - \frac{\partial}{\partial \bar z}.
\end{equation}
We can define the chiral superfield in such theory as
\begin{equation}
 \overline D_\pm \Phi = 0
\end{equation}
where
\begin{equation}
 \overline D_+ = \frac{1}{\sqrt 2}\left(
 \frac{\partial}{\partial \theta_-} + \bar \theta_- \frac{\partial}{\partial z}
\right)
 ,\quad
 \overline D_- = \frac{1}{\sqrt 2}\left(
 \frac{\partial}{\partial \bar \theta_+} + \theta_+
 \frac{\partial}{\partial \bar z}
\right).
\end{equation}
The form of the echiral superfield is
\begin{gather}
 \Phi = \phi(z+\bar\theta_-\theta_-, \bar z - \bar\theta_+\theta_+)\nonumber\\
 +\sqrt 2 \bar\theta_- \psi_+(z+\bar\theta_-\theta_-, \bar z - \bar\theta_+\theta_+)
 +\sqrt 2 \theta_+ \bar\psi_-(z+\bar\theta_-\theta_-, \bar z -
 \bar\theta_+\theta_+) \nonumber\\
 + 2\theta_+\bar\theta_- D(z+\bar\theta_-\theta_-, \bar z - \bar\theta_+\theta_+)
\end{gather}
where $\psi_\pm$ and $\bar\psi_\pm$ are chiral components of the Dirac
fermion
\begin{equation}
 \psi_\pm = \frac{1\pm\gamma_5}{2}\psi , \quad
 \bar\psi_\mp = \bar\psi \frac{1\pm\gamma_5}{2}.
\end{equation}
On the other hand, anti-chiral superfield is written as
\begin{gather}
 \bar\Phi = \phi^* 
(z-\bar\theta_-\theta_-, \bar z + \bar\theta_+\theta_+)\nonumber\\
 -\sqrt 2 \theta_- \bar\psi_+(z-\bar\theta_-\theta_-, \bar z + \bar\theta_+\theta_+)
 -\sqrt 2 \bar \theta_+ \psi_-(z-\bar\theta_-\theta_-, \bar z +
 \bar\theta_+\theta_+) \nonumber\\
 - 2\bar\theta_+\theta_- D^*(z-\bar\theta_-\theta_-, \bar z +
 \bar\theta_+\theta_+)
 .
\end{gather}
By calculating $\epsilon Q_+ \Phi$, $\epsilon Q_+ \bar \Phi$, we
introduce supersymmetry transformation as
\begin{alignat}{2}
 &\delta_1 \phi = \epsilon \psi_+ , &\qquad
 &\delta_1 \phi^* = 0 , \nonumber\\
 &\delta_1 \psi_+ = 0 , &\qquad
 &\delta_1 \psi_- = 0 , \nonumber\\
 &\delta_1 \bar\psi_-  = \epsilon D , &\qquad
 &\delta_1 \bar\psi_+ = - \epsilon \partial_z \phi^* , \nonumber\\
 &\delta_1 D = 0 , &\qquad
 &\delta_1 D^* = \epsilon \partial_z \psi_- ,
\end{alignat}
and from $\epsilon \overline Q_- \Phi$, $\epsilon
\overline Q_- \bar\Phi$, introduce another one
\begin{alignat}{2}
 &\delta_2 \phi = 0 , &\qquad
 &\delta_2 \phi^* = - \epsilon \psi_- , \nonumber\\
 &\delta_2 \psi_+ = 0 , &\qquad
 &\delta_2 \psi_- = 0 , \nonumber\\
 &\delta_2 \bar\psi_-  = \epsilon \partial_{\bar z} \phi , &\qquad
 &\delta_2 \bar\psi_+ = -\epsilon D^* , \nonumber\\
 &\delta_2 D = - \epsilon \partial_{\bar z} \psi_+ , &\qquad
 &\delta_2 D^* = 0 .
\end{alignat}
Now we take Lagrangian ${\cal L}$ as
\begin{gather}
 [\bar\Phi \Phi]_{\mbox{\scriptsize D-term}} +
 \left([W(\Phi)]_{\mbox{\scriptsize F-term}}+ h.c.\right) \nonumber\\
 \cong \bar\psi_+ \partial_{\bar z} \psi_+ + \bar\psi_- \partial_{z} \psi_-
 + \partial_z \phi^* \partial_{\bar z} \phi - D^* D \nonumber\\
 + \bar\psi_- W'' \psi_+      - W' D
 + \bar\psi_+ W^*{}'' \psi_-  - W^*{}' D^* \nonumber\\
 \equiv {\cal L},
\end{gather}
where we have arranged total divergence terms appropriately.
The variation of this Lagrangian under $\delta_1$ gives
total divergence term
\begin{equation}
 -\epsilon (\psi_- \partial_z W^*{}' + W^*{}'\partial_z \psi_-).
\end{equation}
On the other hand, the variation of $W^*{}'\partial_z \phi^*$
under $-\delta_2$ gives
\begin{align}
 -\delta_2 ( W^*{}'\partial_z \phi^*) & =
 - W^*{}' \partial_z(-\epsilon \psi_-) - (-\epsilon \psi_-) W^*{}''
 \partial_z \phi^*
 \nonumber\\
 & = \epsilon (\psi_- \partial_z W^*{}' + W^*{}'\partial_z \psi_-).
\end{align}
So if we redefine Lagrangian including
$W^*{}'\partial_z \phi^*$ and its complex conjugate
\begin{equation}
 \tilde{\cal L} \equiv
 {\cal L} + W' \partial_{\bar z}\phi + W^*{}'\partial_z \phi^*
 , \label{eq:susy-lagrangian}
\end{equation}
then we have the symmetry under $\delta_1 - \delta_2$ at the
Lagrangian level.

Now let us see the relation between this symmetry and the Nicolai mapping.
The Nicolai mapping in continuum is written as
\begin{gather}
 M = \partial_1 A - \partial_2 B + U \\
 N = -\partial_2 A - \partial_1 B + V
\end{gather}
so that fermionic action is given by
\begin{equation}
\bar\psi (D+F) \psi
  = \left( \bar\psi_1 \ \bar\psi_2 \right)
\left(\begin{matrix}
  \frac{\partial M}{\partial A} &
  -i\frac{\partial M}{\partial B} \\
  i\frac{\partial N}{\partial A} &
  \frac{\partial N}{\partial B}
\end{matrix}\right)
\left(\begin{matrix}
\displaystyle \psi_1 \\
\displaystyle \psi_2
\end{matrix}\right)
\end{equation}
where $A$ and $B$ are real and imaginary part (normalized by
$1/\sqrt{2}$) of $\phi$ and $U$ and $V$ are those of $W'$.
The total action implied by the Nicolai mapping is just equal to
(\ref{eq:susy-lagrangian}). We can also see that the supersymmetry
transformation implied by the Nicolai mapping
\begin{alignat}{2}
 &\delta A = \epsilon \psi_1 , &\quad
 &\delta B = - i \epsilon \psi_2, \nonumber\\
 &\delta \psi_1 = 0 , &\quad
 &\delta \psi_2 = 0, \nonumber\\
 &\delta \bar \psi_1 = - \epsilon M , &\quad
 &\delta \bar \psi_2 = i \epsilon N
 \label{eq:susy-transformation-delta}
\end{alignat}
is nothing but the transformation $\delta_1 - \delta_2$ described above.
Actually, by inserting
\begin{alignat}{2}
 &\psi_+ = \frac{1}{\sqrt 2}(\psi_1 + \psi_2) , &\quad
 &\psi_- = \frac{1}{\sqrt 2}(\psi_1 - \psi_2) \nonumber\\
 &\bar\psi_+ = \frac{1}{\sqrt 2}(\bar\psi_1 - \bar\psi_2) , &\quad
 &\bar\psi_- = \frac{1}{\sqrt 2}(\bar\psi_1 + \bar\psi_2) \\
\end{alignat}
into (\ref{eq:susy-transformation-delta}), then we obtain
\begin{alignat}{2}
 &\delta \phi = \epsilon \psi_+ , &\quad
 &\delta \phi^* = \epsilon \psi_- \nonumber\\
 &\delta \bar \psi_- = -\epsilon \partial_{\bar z}\phi -\epsilon W^*{}' , &\quad
 &\delta \bar \psi_+ = -\epsilon \partial_{z}\phi^* -\epsilon W'
\end{alignat}
and this coincides with $\delta_1 - \delta_2$ after
eliminating auxiliary fields $D$ and $D^*$ by their equation of motion.


\begin{thebibliography}{9}

\bibitem{Ginsparg-Wilson}
P.H.~Ginsparg and K.G.~Wilson,
Phys. Rev. D25 (1982) 2649.

\bibitem{PH-fp-action}
P.~Hasenfratz,
Nucl. Phys. B(Proc. Suppl.)63 (1998) 53.

\bibitem{HN-chiralsym}
H.~Neuberger,
Phys. Lett. B417 (1998) 141;
Phys. Lett. B427 (1998) 353.

\bibitem{ML-chiralsym}
M.~L\"uscher,
Phys. Lett. B428 (1998) 342.

\bibitem{Fujikawa-algebraic-extension}
K.~Fujikawa,
Nucl. Phys. B589 (2000) 487;
K.~Fujikawa and M.~Ishibashi,
Nucl. Phys. B587 (2000) 419;
Nucl. Phys. B605 (2001) 365.

\bibitem{staggered-susy}
T.~Banks and P.~Windey, Nucl. Phys. B198 (1982) 226;
S.~Elitzur, E.~Rabinovici and A.~Schwimmer, Phys. Lett. 119B (1982) 165; 
R.~Nakayama and Y.~Okada, Phys. Lett. 134B (1984) 241; 
I.~Ichinose, Phys. Lett. 122B (1983) 68.

\bibitem{slac-susy} 
J.~Bartels and J.B.~Bronzan, 
Phys. Rev. D28 (1983) 818.

\bibitem{nojiri}
S.~Nojiri, Prog. Theor. Phys. 74 (1985) 819.

\bibitem{curci-veneziano}
G.~Curci and G.~Veneziano, Nucl. Phys. B292 (1987) 555.

\bibitem{wilson-susy}
J.~Bartels and G.~Kramer, Z. Phys. C20 (1983) 159.

\bibitem{taniguchi-WT-Wilson}
Y.~Taniguchi,
Phys. Rev. D63 (2001) 014502.

\bibitem{dwf}
D.B.~Kaplan, Phys. Lett. B288 (1992) 342.

\bibitem{dwf-vector}
Y.~Shamir, Nucl. Phys. B406 (1993) 90; 
V.~Furman and Y.~Shamir, Nucl. Phys. B439 (199) 54.

\bibitem{overlap}
R.~Narayanan and H.~Neuberger,
Nucl. Phys. B412 (1994) 574;
Phys. Rev. Lett. 71 (1993) 3251;
Nucl. Phys. B443 (1995) 305.

\bibitem{nishimura-SYM}
J.~Nishimura,
Phys. Lett. B406 (1997) 215; 
%
N.~Maru and J.~Nishimura,
Int. J. Mod. Phys. A13 (1998) 2841.


\bibitem{neuberger-SYM}
H.~Neuberger,
Phys. Rev. D57 (1998) 5417.

\bibitem{aoyama-kikukawa}
T.~Aoyama and Y.~Kikukawa,
Phys. Rev. D59 (1999) 054507.


\bibitem{kaplan-shmaltz}
D.B.~Kaplan and M.~Schmaltz,
Chin. J. Phys. 38 (2000) 543.

\bibitem{bietenholz-susy}
W.~Bietenholz,
Mod. Phys. Lett. A14 (1999) 51.

\bibitem{so-ukita-susy}
H.~So and N.~Ukita
Phys. Lett. B457 (1999) 314.

\bibitem{KF-MI-wz}
K.~Fujikawa and M.~Ishibashi,
Nucl. Phys. B622 (2002) 115.

\bibitem{Fujikawa-SUSY-Leibniz}
K.~Fujikawa,
Nucl. Phys. B636 (2002) 80.

\bibitem{Nicolai-mapping}
H.~Nicolai,
Phys. Lett. 89B (1980) 341.

\bibitem{spacial-lattice-nicolai}
S.~Cecotti and L.~Girardello, Nucl. Phys. B226 (1983) 417.

\bibitem{NS-MS-nicolai}
N.~Sakai and M.~Sakamoto,
Nucl. Phys. B229 (1983) 173.

\bibitem{so-kato-susy}
K.~Itoh, M.~Kato, H.~Sawanaka, H.~So and N.~Ukita,
{\tt hep-lat/0112052}.

\bibitem{kaplan-susy-spatial-lattice}
D.B.~Kaplan, E.~Katz and M.~Unsal,
{\tt hep-lat/0206019}.

\bibitem{SC-SK-susy}
S.~Catterall and S.~Karamov,
Phys. Rev. D65 (2002) 094501.

\bibitem{Parisi-Sourlas}
G.~Parisi and N.~Sourlas,
Nucl. Phys. B206 (1982) 321.

\bibitem{KF-MI-majorana}
K.~Fujikawa and M.~Ishibashi,
Phys. Lett. B528 (2002) 295.

\end{thebibliography}
\end{document}